\documentclass[preprint]{elsarticle}
\usepackage{amsmath}
\usepackage{amssymb}
\usepackage{lipsum}
\usepackage{gensymb}
\usepackage[usenames,dvipsnames]{xcolor}
\usepackage{graphicx}% Include figure files
\usepackage{dcolumn}% Align table columns on decimal point
\usepackage{bm}% bold math
\usepackage{comment}
\usepackage{siunitx}
\usepackage[numbers]{natbib}
\DeclareSIUnit{\wtwtpercent}{wt/wt\%}
\DeclareSIUnit\Molar{\textsc{m}}
\DeclareSIUnit\molar{\textsc{m}}

\journal{}

\makeatletter
\def\ps@pprintTitle{%
 \let\@oddhead\@empty
 \let\@evenhead\@empty
 \let\@oddfoot\@empty
 \let\@evenfoot\@empty}
\makeatother

\begin{document}
\begin{frontmatter}

\title{Anisotropic Diffusion in Lyotropic Chromonic Liquid Crystal using Fluorescence Recovery After Photobleaching}

\author[1]{Kyu Hwan Choi}

\author[2]{Jiyong Cheon}

\author[3,4]{Joonwoo Jeong\corref{cor1}}
\ead{jjeong@unist.ac.kr}

\author[1]{Sho C. Takatori\corref{cor1}}
\ead{stakatori@stanford.edu}
\cortext[cor1]{Corresponding author}

% ---- Affiliations ----

\affiliation[1]{organization={Department of Chemical Engineering, Stanford University},
            city={Stanford},
            state={CA},
            country={USA}}

\affiliation[2]{organization={School of Physics, College of Sciences, Georgia Institute of Technology},
            city={Atlanta},
            state={GA},
            country={USA}}

\affiliation[3]{organization={Department of Physics, Ulsan National Institute of Science and Technology},
            city={Ulsan},
            country={Republic of Korea}}

\affiliation[4]{organization={UNIST Research Center for Soft and Living Matter, Ulsan National Institute of Science and Technology},
            city={Ulsan},
            country={Republic of Korea}}

\begin{abstract}
Anisotropic diffusion governs transport in a wide range of soft and biological materials, where microstructure and molecular interactions jointly shape how matter moves. Here, we quantitatively investigate anisotropic molecular transport in lyotropic chromonic liquid crystals (LCLCs) using fluorescence recovery after photobleaching (FRAP). Disodium cromoglycate (DSCG) serves as a model LCLC system, and diffusion is measured across isotropic, nematic, and columnar phases as concentration and temperature are varied.

To disentangle the roles of microstructure and molecular interactions, we employ two fluorescent tracers with distinct affinities for the LCLC aggregates: Acridine Orange (AO), which intercalates into DSCG aggregates, and Bodipy, which interacts weakly and remains largely in the aqueous phase. Fourier-space FRAP analysis independently resolves the parallel and perpendicular diffusion coefficients for both dyes relative to the liquid-crystal alignment.

In the nematic phase, diffusion becomes anisotropic, with faster transport along the liquid-crystal director. As the DSCG concentration increases, AO dye molecules that are strongly coupled to the aggregates exhibit a slowdown in all directions, reflecting enhanced packing and steric confinement of the LC microstructure. In contrast, weakly interacting Bodipy dye molecules display enhanced transport along the alignment direction as the DSCG concentration increases in the nematic regime, suggesting the emergence of microscopic channels that guide motion, analogous to transport in oriented porous media.
These results reveal how the evolving microstructure of LCLCs controls effective diffusion and provide a quantitative framework for understanding and designing anisotropic transport in aligned soft materials.
\end{abstract}

\begin{keyword}
liquid crystal \sep anisotropic diffusion \sep diffusion mechanism
\end{keyword}

\end{frontmatter}

\section{Introduction}
\label{introduction}
Macromolecules in biological systems often exhibit anisotropic diffusion due to the order and structure of their surrounding environment. For example, in mammalian cell membranes, diffusion occurs at different rates in-plane and out-of-plane of the lipid bilayer\cite{travascio2009characterization,smith1979anisotropic}. Understanding anisotropic diffusion is crucial for elucidating material transport in such systems, particularly in liquid crystalline environments that mimic biological structures such as membranes and extracellular matrices\cite{smith1979anisotropic, chen2021noninvasive}.

In lyotropic chromonic liquid crystals (LCLCs), plank-like molecules self-assemble to make rod-like aggregates via face-to-face stacking interactions\cite{Nastishin2005, lydon2011chromonic, lydon1998chromonic, eun2020lyotropic}.
At concentrations above a certain threshold, the aggregates align to form liquid crystalline phases: nematic and columnar phases. 
In this respect, LCLCs share key physical features with biological materials such as DNA bundles, collagen matrices, and lipid bilayers: 
they develop microscopic structural alignment that gives rise to direction-dependent transport.
In all of these systems, molecular motion is guided by an underlying anisotropic architecture, leading to faster transport along preferred directions and hindered motion across them\cite{leddy2006diffusional,lucchetti2020elasticity,kim2015lyotropic}.
Despite their structural and transport relevance to biological systems, 
how anisotropic diffusion in LCLCs emerges from the interplay between microscopic alignment and molecular-scale interactions remains poorly understood. 
Previous studies have largely focused on either the self-diffusion of lyotropic liquid crystals or the diffusion of dissolved molecules, without simultaneously characterizing both within the same structured system\cite{leddy2006diffusional,lettinga2005self,allen1990diffusion,chen2021noninvasive}.

To address this gap, we investigate anisotropic diffusion in lyotropic chromonic liquid crystals using disodium cromoglycate (DSCG) as a model system and fluorescence recovery after photobleaching (FRAP) as a quantitative measurement method. Two fluorescent dyes with distinct strengths of interactions with LCLC aggregates are employed to disentangle the diffusion of DSCG aggregates from that of molecular tracers in the surrounding medium. By systematically varying the DSCG concentration, we examine molecular transport across the isotropic, nematic (N) and columnar (M) phases.

FRAP provides a robust measure of bulk effective diffusivity by monitoring the spatiotemporal recovery of fluorescence following localized photobleaching\cite{mullineaux2007using,axelrod1976mobility}. In this study, the interpretation of FRAP measurements is guided by independent characterization of dye--DSCG molecule interactions using UV--Vis spectroscopy\cite{yang2022orientational}, allowing diffusional behavior to be directly linked to molecular localization within the LCLC system. To quantify transport anisotropy, fluorescence recovery is analyzed in Fourier space, enabling independent extraction of diffusion coefficients parallel and perpendicular to the liquid-crystal alignment\cite{chen2021noninvasive,travascio2007anisotropic,smith1979anisotropic,tsay1991spatial,berk1993fluorescence,travascio2009characterization}. Together, this framework provides a reliable and quantitative link between liquid-crystal microstructure and anisotropic molecular transport in LCLCs.

\section{Experimental Methods}

\subsection{LCLC Preparation}
Dye-contained DSCG aqueous solution was prepared for FRAP measurements by dissolving DSCG ($\geq 95 \%$, Sigma-Aldrich) and fluorescent dyes in DI-water ($\geq$ \SI{18.2}{\mega\ohm\cdot\centi\meter}) at DSCG concentrations ranging from \SI{16}{} to \SI{20}{\wtwtpercent}. Two different dyes, Acridine Orange (AO) and Bodipy-NHS, were obtained from Thermo Fisher and added at final concentrations of \SI{3}{\micro\Molar} and \SI{50}{\nano\Molar}, respectively.
These low dye concentrations, \SI{0.05}{}-\SI{3}{\micro\Molar}, minimize disturbance to the pristine liquid crystals' microstructure\cite{yang2022orientational}.
No additional purification steps were performed for any chemicals used in this study.

\subsection{Fabrication of Observation Cell}
The FRAP measurement was conducted using a thin-gap sandwich cell, consisting of a larger coverslip (\SI{50}{\milli\meter} $\times$ \SI{25}{\milli\meter}), a smaller coverslip (\SI{22}{\milli\meter} $\times$ \SI{22}{\milli\meter}), and a confined LCLC solution. The liquid crystal was aligned by mechanical rubbing, with both glass substrates rubbed in a single direction using abrasive pads (Trizact 3000, 3M). The rubbed glass substrates were sonicated for \SI{30}{\minute} in ethanol (Sigma-Aldrich) and DI water, followed by nitrogen drying.

The cell thickness, ranging from \SI{5}{} to \SI{7.5}{\micro\meter}, was controlled by adjusting the volume of the liquid crystal solution (\SI{2.5}{}-\SI{3}{\micro\liter}) deposited onto the substrate. The coverslip was gently placed to spread the droplet, ensuring the abrasive direction remained aligned. To minimize the water evaporation, the coverslip edges were sealed with epoxy glue (Devcon). To eliminate shear-flow effects and thermal history, the assembled samples were heated to \SI{60}{\celsius} for \SI{30}{\minute}, during which the DSCG solutions exhibited an isotropic phase.

For columnar phase solutions at \SI{20}{\wtwtpercent} DSCG, all components, including glass substrates, pipette tips, and the solution itself, were preheated to \SI{50}{\celsius} to facilitate handling of the columnar phase solution by reducing its viscosity. Since water evaporates rapidly at high temperatures, sample preparation was completed within one minute after depositing the solution.

LCLC's phase and homogeneous alignment were verified using a polarized optical microscope (Ti-2 Eclipse, Nikon). The alignment was examined at both \SI{0}{\degree} (parallel) and \SI{90}{\degree} (crossed) angles between the polarizer and analyzer.

\subsection{Fluorescence Recovery After Photobleaching (FRAP)}
FRAP measurements were performed using an inverted fluorescence microscope equipped with a water-immersion 60$\times$ objective lens (CFI plan Apo VC, Numerical Aperture 1.2, Nikon). Multi-Line LED Light Source (SpectraX, Lumencor) was used for fluorescence excitation, with wavelengths of \SI{488}{\nano\meter} for Acridine Orange (AO) and Bodipy. The temperature was controlled by the heating stage (H301-Mini, OKOlab). % 647nm for Cy5

\begin{figure}
	\centering
	\includegraphics[width=0.9\columnwidth]{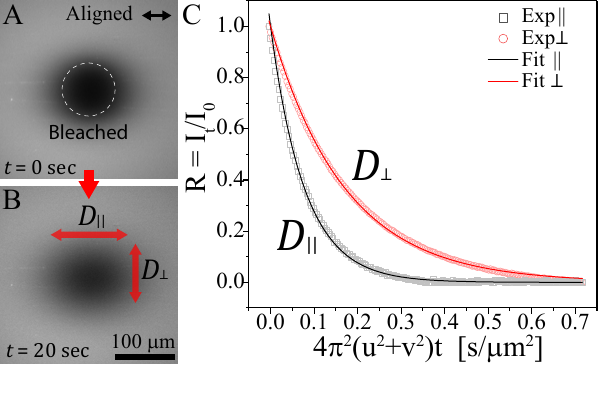}
	\caption{FRAP measurements of anisotropic diffusion. A) Fluorescence image taken immediately after photobleaching ($t =$ \SI{0}{\second}) in \SI{18}{\wtwtpercent} DSCG with acridine orange (AO); the dashed line indicates the bleached region. B) Fluorescence image at $t =$ \SI{20}{\second} during recovery. C) Time-dependent intensity decay of single Fourier modes used to extract the diffusion coefficients, the mode at $(u, v) = (0.014, 0)$\SI{}{\per\micro\meter} for the parallel component $D_{\parallel}$, and the mode at $(u, v) = (0, 0.014)$\SI{}{\per\micro\meter} for the perpendicular component $D_{\perp}$. Open symbols are experimental data, solid lines are fits to Eq.~\ref{eq : fit}. Black and red correspond to parallel ($D_\parallel$) and perpendicular ($D_\perp$) diffusion, respectively.}
	\label{fig:fourier}
\end{figure}

The bleaching spot size was controlled using a circular physical iris with an approximate diameter of \SI{70}{}-\SI{100}{\micro\meter}, as indicated by the white dashed circle in Fig.~\ref{fig:fourier}A. The laser was applied at full power (\SI{100}{\percent}, \SI{196}{\milli\watt} at the laser head before the optical path) for \SI{60}{\second} to bleach the designated region. Immediately after bleaching, the recovery process was recorded with the laser power reduced to \SI{3}{\percent}, to minimize further photobleaching.

Fluorescence recovery images were recorded using a CMOS camera (Photometrics Prime 95B, Teledyne Photometrics) for a total duration of \SI{120}{} or \SI{240}{\second}, with frames captured at \SI{0.5}{} or \SI{1}{\second} intervals.(Fig.~\ref{fig:fourier}B) The entire FRAP process, including bleaching and recovery imaging, was controlled through Micromanager on ImageJ.

\subsection{Diffusion Coefficient Extraction from FRAP}
To extract diffusion coefficients in the parallel ($D_{||}$) and perpendicular ($D_\perp$) directions, a Fourier transform-based analysis was applied to FRAP image sequences\cite{chen2021noninvasive,travascio2007anisotropic, smith1979anisotropic, tsay1991spatial}. Background intensity variations were removed by normalizing all images with respect to the equilibrium image after recovery, $I_t(x,y,t)=i(x,y,t)/i_0(x,y)$, where $I_t(x,y,t)$, $i(x,y,t)$, and $i_0(x,y)$ are normalized intensity, raw intensity at specific time $t$, and raw intensity before bleaching, respectively (Fig.~\ref{fig:fourier}A).

After normalization, the images were Fourier-transformed, $\mathcal{F}(I_t(x,y))=\hat{I_t}(u,v)$, and the second or third points along the $u$ and $v$ axes were selected for analysis in Fourier space, which corresponds to \SI{0.014}{}-\SI{0.045}{\per\micro\meter} (Fig.~\ref{fig:fourier}B). These specific frequency points were chosen because lower frequencies correspond to large-scale diffusion dynamics, \SI{23}{}-\SI{72}{\micro\meter}.

The decay of time-dependent relative intensity, $R(u,v,t)=\hat{I_t}(u,v,t)/\hat{I}(u,v,0)$, was fitted using the following equations (Fig.~\ref{fig:fourier}C):
\begin{equation}
R_{||} = \exp(-4\boldsymbol{D_{||}}\pi^2 u^2 t), \quad
R_{\perp} = \exp(-4\boldsymbol{D_{\perp}}\pi^2 v^2 t).
\label{eq : fit}
\end{equation}

The horizontal axis, $4\pi^2 (u^2+v^2)t$ [\si{\second\per\micro\meter^2}], represents a diffusion time associated with a given spatial frequency $(u,v)$, such that the decay directly reflects diffusive relaxation at the corresponding length scale. 

The diffusion coefficients were obtained by fitting the relative intensity decay profiles to Eq.~\ref{eq : fit}. Additional details regarding the analysis procedure can be found in the Supporting Information (SI).

\subsection{UV--Vis Spectroscopy for Dye Interaction}
\begin{figure}
    \centering
    \includegraphics[width=\columnwidth]{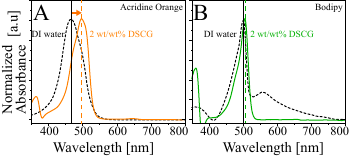} 
    \caption{Normalized UV-vis absorption spectra of two dyes dissolved in two different media: DI-water and \SI{2}{\wtwtpercent} DSCG solution. 
    A) Acridine Orange. The peak at \SI{465}{\nano\meter} in DI-water shifted to \SI{500}{\nano\meter} in \SI{2}{\wtwtpercent} DSCG solution. B) Bodipy. No shift in the peak wavelength was observed. 
    Black dash lines represent the UV-vis spectrum of dye dissolved in DI-water, and colored lines correspond to the \SI{2}{\wtwtpercent} DSCG aqueous solution.
    }
    \label{fig:UV-vis}
\end{figure}

The interaction strengths between fluorescent dyes and DSCG liquid-crystal aggregates were characterized using UV--Vis absorption spectroscopy (NanoDrop 2000C, Thermo Fisher Scientific)\cite{yang2022orientational} (Fig.~\ref{fig:UV-vis}). Spectra were acquired for dyes dissolved in two different media: deionized (DI) water and a \SI{2}{\wtwtpercent} DSCG aqueous solution, enabling direct comparison
of how dye spectra change because of the interaction with DSCG aggregates, e.g., peak shifts.

Acridine Orange (AO) and Bodipy were prepared at concentrations of \SI{1}{\milli\molar} and \SI{0.1}{\milli\molar}, respectively, which are sufficient to ensure reliable absorbance signals while avoiding aggregation or perturbation of the LC structure\cite{mundy1995intercalation,yang2022orientational}. All absorbance spectra were normalized by the maximum intensity measured in the \SI{2}{\wtwtpercent} DSCG solution. The wavelength corresponding to the maximum absorbance was identified as the peak wavelength and used to assess dye–LC interactions.

\section{Results and Discussion}
\subsection{Dye--Aggregate Interaction Strength}
The interaction strengths between the fluorescent dyes and DSCG aggregates were determined using UV-Vis spectroscopy, as represented in Fig.~\ref{fig:UV-vis}. A shift in the maximum absorbance peak is interpreted as evidence of strong dye–DSCG molecule interactions, such as intercalation within DSCG stacks. In contrast, the absence of a peak shift hints at weak or negligible interaction and free diffusion in the aqueous medium\cite{yang2022orientational}.

The absorbance data provide direct insight into dye localization within the LCLC system. Acridine Orange (AO) exhibits a pronounced peak shift from \SI{465}{\nano\meter} to \SI{500}{\nano\meter} upon addition of DSCG (Fig.~\ref{fig:UV-vis}A), indicating strong interaction with the LCLC. This behavior is consistent with previous reports that the planar molecular structure of AO intercalates into planar chromonic stacks\cite{yang2022orientational}. In contrast, Bodipy shows no measurable shift in its absorbance peak (Fig.~\ref{fig:UV-vis}B), indicating weak interaction with the LCLC molecules. Accordingly, the diffusion of AO reflects the motion of DSCG aggregates, whereas Bodipy remains largely dissolved in the aqueous medium and serves as a probe of free molecular diffusion within the LCLC system.

These peak shifts and interaction strengths are consistent with the molecular structures and electrostatic properties of the dyes. Negatively charged LC molecules interact more strongly with the positively charged AO, while neutral Bodipy remains freely dissolved due to weaker interactions with the LCLC molecules.

\subsection{Effect of DSCG Concentration on LCLC Structure}
The LCLC microstructures depend on concentration and temperature, as analyzed via X-ray scattering~\cite{horowitz2005aggregation, cheon2025solvent}. The results, described in Fig.~S1, show that the inter-aggregate distance, $d\sim$\SI{4.99}{}-\SI{4.25}{\nano\meter}, and the inter-molecular correlation length, $\xi_\parallel\sim$\SI{7.77}{}-\SI{8.56}{\nano\meter}, exhibit relatively weak dependence on the increasing DSCG concentration from $\sim$\SI{12.5}{\wtwtpercent} to $\sim$\SI{20}{\wtwtpercent}. In contrast, the inter-aggregate correlation length, $\xi_\perp$, increases significantly from \SIrange{6.99}{20.04}{\nano\meter} in the same concentration range. 
Higher DSCG concentration results in effectively longer LC aggregates with longer inter-aggregate correlation length, $\xi_\perp$, and higher packing density. These structural insights are crucial in explaining the concentration-dependent diffusion behavior discussed below.

\begin{figure}
    \centering
    \includegraphics[width=0.9\columnwidth]{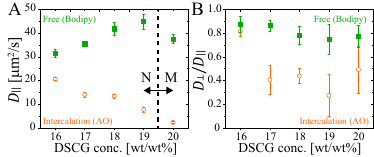} 
    \caption{
    Diffusivities of Acridine Orange (AO, orange) and Bodipy (green) as a function of DSCG concentration. 
    A) Diffusion coefficients along the aligned director measured at a constant temperature of \SI{25}{\celsius} for AO (orange) and Bodipy (green).  
    %Filled symbols represent diffusion parallel to the nematic director ($D_{\parallel}$), and open symbols represent diffusion perpendicular to the director ($D_{\perp}$).
    B) Diffusion anisotropy reported as a ratio of diffusivities perpendicular and parallel to the director, $D_{\perp}/D_{\parallel}$.
    }
    \label{fig:concentration}
\end{figure}

\subsection{Effect of DSCG Concentration on Anisotropic Transport} 
Diffusion behavior at a constant temperature (\SI{25}{\celsius}) was evaluated as a function of DSCG concentration using diffusion coefficients measured parallel ($D_{\parallel}$) and perpendicular ($D_{\perp}$) to the LC alignment via FRAP. To quantify deviations from isotropic diffusion, we also analyze the ratio $D_{\perp}/D_{\parallel}$, where $D_{\perp}/D_{\parallel}=1$ corresponds to isotropic diffusion. As shown in Fig.~S2, FRAP measurements performed in the isotropic phase (\SI{16}{\wtwtpercent} at 40-\SI{45}{\celsius}) yield $D_{\perp}/D_{\parallel}\approx 1$ for both dyes. This observation is consistent with previous studies showing that, in the absence of microstructural alignment, diffusion in isotropic soft-matter systems exhibits no directional preference~\cite{dobrindt2012anisotropic}. $D_{\perp}/D_{\parallel} < 1$ indicates that the diffusion along the alignment direction is faster than the diffusion perpendicular to the alignment direction.

For Acridine Orange (AO), which strongly intercalates within DSCG aggregates \cite{yang2022orientational} and therefore reports the self-diffusion of the aggregates, both $D_{\parallel}$(\SIrange{20.66}{2.34}{\micro\meter\squared\per\second}) and $D_{\perp}$ (\SIrange{16.81}{1.15}{\micro\meter\squared\per\second}) decrease with increasing DSCG concentration.
The increasing anisotropy, consequently, reflected in the systematic decrease of $D_{\perp}/D_{\parallel}$ from 0.81~$\pm$~0.04 at \SI{16}{\wtwtpercent} to 0.49~$\pm$~0.20 at \SI{20}{\wtwtpercent} indicates that although diffusion slows in both directions as the LC becomes denser and more ordered, transport perpendicular to the director is suppressed more strongly. 
This enhanced anisotropy is consistent with the pronounced growth of the perpendicular inter-aggregate correlation length of DSCG ($\xi_\perp \sim$ \SI{7}{}–\SI{20}{\nano\meter}).
Such increased lateral correlation would be expected to hinder transverse motion more strongly than motion along the director.

In contrast, the microstructural length scales governing transport along the director, the inter-molecular correlation length ($\xi_\parallel \sim$ \SI{8}{\nano\meter}) and the inter-aggregate spacing ($d \sim$ \SI{4.5}{\nano\meter}), remain nearly constant over this concentration range (Fig.~S1). 
Nevertheless, diffusion along the director still decreases with concentration. We hypothesize that increasing lateral correlations may enhance transverse confinement and inter-aggregate interactions, which could effectively increase resistance to longitudinal motion.
Consequently, although both $D_{\parallel}$ and $D_{\perp}$ decrease with increasing concentration, transverse transport is suppressed more strongly, leading to a systematic decrease in the anisotropy ratio $D_{\perp}/D_{\parallel}$.
Overall, the decrease in $D_{\perp}/D_{\parallel}$ reflects the progressive structural ordering of the LCLC system at higher concentrations, approaching the columnar regime, where transverse motion becomes increasingly constrained~\cite{allen1990diffusion,lettinga2005self,alavi1992anomalous}.

Overall, the decrease in $D_{\perp}/D_{\parallel}$ reflects the progressive structural ordering of the LCLC system at higher concentrations, approaching the columnar regime, where transverse motion becomes increasingly constrained~\cite{allen1990diffusion,lettinga2005self,alavi1992anomalous}.

In contrast, Bodipy, which diffuses primarily through the aqueous medium and interacts weakly with DSCG, displays relatively isotropic diffusion with $D_{\perp}/D_{\parallel} \approx$ 0.75-0.87 compared to diffusion of AO across the experimental range (\SIrange{16}{20}{\wtwtpercent}). 
Notably, $D_{\parallel}$ increases with concentration (\SIrange{31.6}{44.9}{\micro\meter\squared\per\second}), suggesting that enhanced lateral correlation promotes the formation of continuous, low-resistance pathways along the director. In such aligned nematic states, the aqueous regions between aggregates become increasingly correlated along the director, enabling more efficient longitudinal transport for weakly interacting tracers. This guided motion is similar to transport in aligned porous media~\cite{kim1987diffusion,quintard1993diffusion,nakashima2010anisotropic}.
However, upon entering the columnar phase at $\geq$ \SI{20}{\wtwtpercent}, both $D_{\parallel}$ and $D_{\perp}$ decrease, consistent with the emergence of a densely packed microstructure that restricts molecular motion in all directions.

The ratio $D_{\perp}/D_{\parallel}$ captures how concentration-driven ordering tunes anisotropic transport. At low concentrations, diffusion is nearly isotropic. At intermediate concentrations, nematic alignment promotes directionally guided motion. At high concentrations, the columnar phase imposes strong confinement. These results highlight the close coupling between LCLC microstructure and the strength of interaction between the dye and DSCG aggregates.

We additionally performed diffusion measurements using Cyanine5 (Cy5) as a probe. Compared to AO, Cy5 exhibited a lower parallel diffusivity, with $D_{\parallel} = $\SI{11.08}{}-\SI{1.82}{\micro\meter^2\per\second}, and a higher degree of isotropy, $D_{\perp}/D_{\parallel} = 0.6$--$0.8$, as shown in Fig.~S3B-C. Cy5 is a positively charged molecule, but it exhibits weaker interactions with DSCG aggregates than AO, as shown by the absence of a significant peak shift in the UV–vis spectrum in DSCG solution (Fig.~S3A). We suspect that the reduced $D_{\parallel}$ of Cy5 may arise from differences in molecular structure, including its extended carbon chains, which could limit its intercalation into DSCG aggregates.

\section{Conclusion}

In this study, we investigated anisotropic diffusion in a lyotropic chromonic liquid crystal (LCLC) system using two fluorescent dyes, Acridine Orange (AO) and Bodipy, as probes with distinct interaction strengths. UV--Vis spectroscopy revealed that AO strongly associates with DSCG aggregates, whereas Bodipy remains weakly interacting and largely dissolved in the aqueous medium. Using FRAP, we quantified how these two classes of tracers respond to concentration-dependent LCLC microstructures. 
Across all conditions except the isotropic phase, diffusion was consistently anisotropic ($D_\parallel>D_\perp$). 
With increasing DSCG concentration, AO parallel diffusivity decreased monotonically, reflecting enhanced steric constraints imposed by denser aggregates. 
In contrast, Bodipy exhibited an increase in parallel diffusivity in the nematic phase as the DSCG concentration increased, followed by a decrease upon entering the columnar phase, revealing a transition from structurally guided transport to strong confinement. 
Together, these results demonstrate that anisotropic transport in LCLCs arises from the interplay among microstructure, phase behavior, and the strength of dye–aggregate interactions.

Beyond the specific LCLC system investigated, this work establishes a robust framework for probing anisotropic transport in structured soft matter. 
By leveraging FRAP with tracers of varying interaction strengths, we can disentangle how distinct molecular species navigate the shared microstructured environment. 
This approach is highly applicable to a wide range of biological and bio-inspired materials, including filament networks, extracellular matrices, and membranous assemblies, where transport is often governed by aligned yet heterogeneous architectures\cite{smith1979anisotropic, chen2021noninvasive, leddy2006diffusional,lucchetti2020elasticity,kim2015lyotropic}.
More broadly, these findings point toward new strategies for engineering directional transport in self-assembled materials, in which phase behavior and microstructure can be tuned to selectively guide or hinder molecular motion.

\section*{Acknowledgements}
KHC and SCT are supported by the National Science Foundation under Grant No.~2440029.
SCT is also supported by the Packard Fellowship in Science. JC and JJ acknowledge the financial support from the National Research Foundation of Korea under Grants No. RS-2024-00345749.
The X-ray scattering experiments were performed at the PLS-II 6D UNIST-PAL Beamline of Pohang Accelerator Laboratory (PAL) in the Republic of Korea (proposal number 2023-2nd-6D-A011).

\bibliographystyle{unsrtnat} 
\bibliography{reference}

\end{document}